\documentclass[reprint,amsmath,amssymb,aps,prapplied]{revtex4-2}

\usepackage{graphicx}
\usepackage{dcolumn}
\usepackage{bm}

\usepackage[utf8]{inputenc}
\usepackage[T1]{fontenc}
\usepackage{mathptmx}

\usepackage[usenames,dvipsnames]{color}

\begin{document}

\title{Design rules for active control of narrowband thermal emission using phase-change materials}

\author{Maxime Giteau}
\author{Mitradeep Sarkar}
\author{Maria Paula Ayala}
\author{Michael T. Enders}
\author{Georgia T. Papadakis}
 \email{georgia.papadakis@icfo.eu}
\affiliation{ICFO-Institut de Ciencies Fotoniques, The Barcelona Institute of Science and Technology, Castelldefels, Barcelona 08860, Spain}

\begin{abstract}
We propose an analytical framework to design actively tunable narrowband thermal emitters at infrared frequencies. 
We exemplify the proposed design rules using phase-change materials (PCM), considering dielectric-to-dielectric PCMs (e.g. GSST) and dielectric-to-metal PCMs (e.g. $\mathrm{VO_2}$). 
Based on these, we numerically illustrate near-unity ON-OFF switching and arbitrarily large spectral shifting between two emission wavelengths, respectively.
The proposed systems are lithography-free and consist of one or several thin emitter layers, a spacer layer which includes the PCM, and a back reflector. Our model applies to normal incidence, though we show that the behavior is essentially angle-independent. The presented formalism is general and can be extended to \textit{any} mechanism that modifies the optical properties of a material, such as electrostatic gating or thermo-optical modulation. 

\end{abstract}

\maketitle

The ability to control the spectrum, direction, and polarization of thermal emission is critical for applications including infrared (IR) sources~\cite{schuller_optical_2009,chen_widely_2020}, thermal camouflage~\cite{li_structured_2018}, radiative cooling~\cite{rephaeli_ultrabroadband_2013,raman_passive_2014} and energy conversion~\cite{liu_taming_2011}.
In particular, the possibility of generating spectrally narrowband IR emission has been the object of a very rich literature~\cite{baranov_nanophotonic_2019,greffet_coherent_2002,celanovic_resonant-cavity_2005,inoue_realization_2015,sakurai_ultranarrow-band_2019,wang_ultranarrow_2020,duan_active_2022,ma_narrowband_2022}. While most architectures involve in-plane patterning and/or a large number of layers to reduce the emission bandwidth,
surface phonon polaritons (SPhPs) offer naturally narrowband resonances owing to their large material quality factors~\cite{caldwell_low-loss_2015}.
In particular, it has been recently shown that few-monolayer SPhP-based emitters used in a Salisbury screen configuration~\cite{salisbury_absorbent_1952,fante_reflection_1988} (a 3-layer structure consisting of an emitter, a dielectric spacer, and a back reflector, forming a Fabry-Perot cavity whose resonance wavelength matches that of the emitter) can achieve strong narrowband emission~\cite{zhao_atomic-scale_2021}.

Another active area of nanophotonics is the active control of optical properties~\cite{picardi_dynamic_2022,fan_active_2022}. Such dynamic modulation can take the form of electrical gating~\cite{de_zoysa_conversion_2012,jang_tunable_2014,inoue_realization_2014,wang_infrared_2015,chen_widely_2020,duan_active_2022}, optical biasing~\cite{chen_experimental_2008} or applied strain~\cite{papadakis_deep-subwavelength_2021}.
In the mid-IR region, the spectral range of interest for thermal emission, a popular approach for active tuning is phase-change materials (PCMs). These materials show a dramatic reversible and (for some of them) non-volatile change in their optical properties upon heating, leading to a very different spectral response~\cite{tittl_switchable_2015,du_control_2017,cao_tuneable_2018,kalantari_osgouei_active_2021,zheng_thermal_2022}.
Several techniques have been developed to induce the phase change beyond simple thermal heating, which is slow and can result in significant hysteresis~\cite{ko_vanadium_2021}. For volatile PCMs, the phase change can be triggered by laser heating, with a characteristic switching time of a few tens of nanoseconds~\cite{ryckman_ultra-compact_2013}. Applying an electrical current can also result in ultrafast switching in a few nanoseconds~\cite{markov_optically_2015}. In the case of non-volatile PCMs, crystallization and amorphization are typically triggered by short (in the order of nanoseconds, depending on the film thickness) laser pulses~\cite{abdollahramezani_tunable_2020}.

Two classes of materials emerge from this description: those switching from one dielectric phase to another, with different refractive indices, such as some GeSbTe (GST) compounds~\cite{michel_reversible_2014,wuttig_phase-change_2017,cao_fundamentals_2019,abdollahramezani_tunable_2020}, and those switching from a dielectric to a metallic phase, such as $\mathrm{VO_2}$~\cite{kats_vanadium_2013,ko_vanadium_2021}. 
PCMs have been studied for various applications including non-volatile optical switching~\cite{michel_reversible_2014,li_reversible_2016,stegmaier_nonvolatile_2017,zhang_broadband_2018,chen_neuromorphic_2022,wan_switchable_2022}, beam switching and bifocal lensing~\cite{yin_beam_2017}, homeostasis~\cite{kats_vanadium_2013}, radiative cooling~\cite{ko_vanadium_2021} and thermal camouflage~\cite{buhara_mid-infrared_2021,king_wavelength-by-wavelength_2022}. They are particularly relevant for spectrally-tunable narrowband sources~\cite{cao_mid-infrared_2013,michel_reversible_2014,chen_tunable_2015,ma_narrowband_2022,xiong_extremely_2022}, with applications in spectroscopy as well as thermophotovoltaics.
However, simple, lithography-free structures tend to have relatively broadband emissivity~\cite{du_control_2017}. Tunable narrowband sources have been achieved only for more complex structures, with an emissivity that is usually not unitary over the whole range of operation~\cite{cao_mid-infrared_2013,xiong_extremely_2022,ma_narrowband_2022}. 
Furthermore, all these devices have limited spectral tunability as they rely on the temperature dependence of a single material's resonance wavelength.

\begin{figure}[h]
    \centering
    \includegraphics[width = \columnwidth]{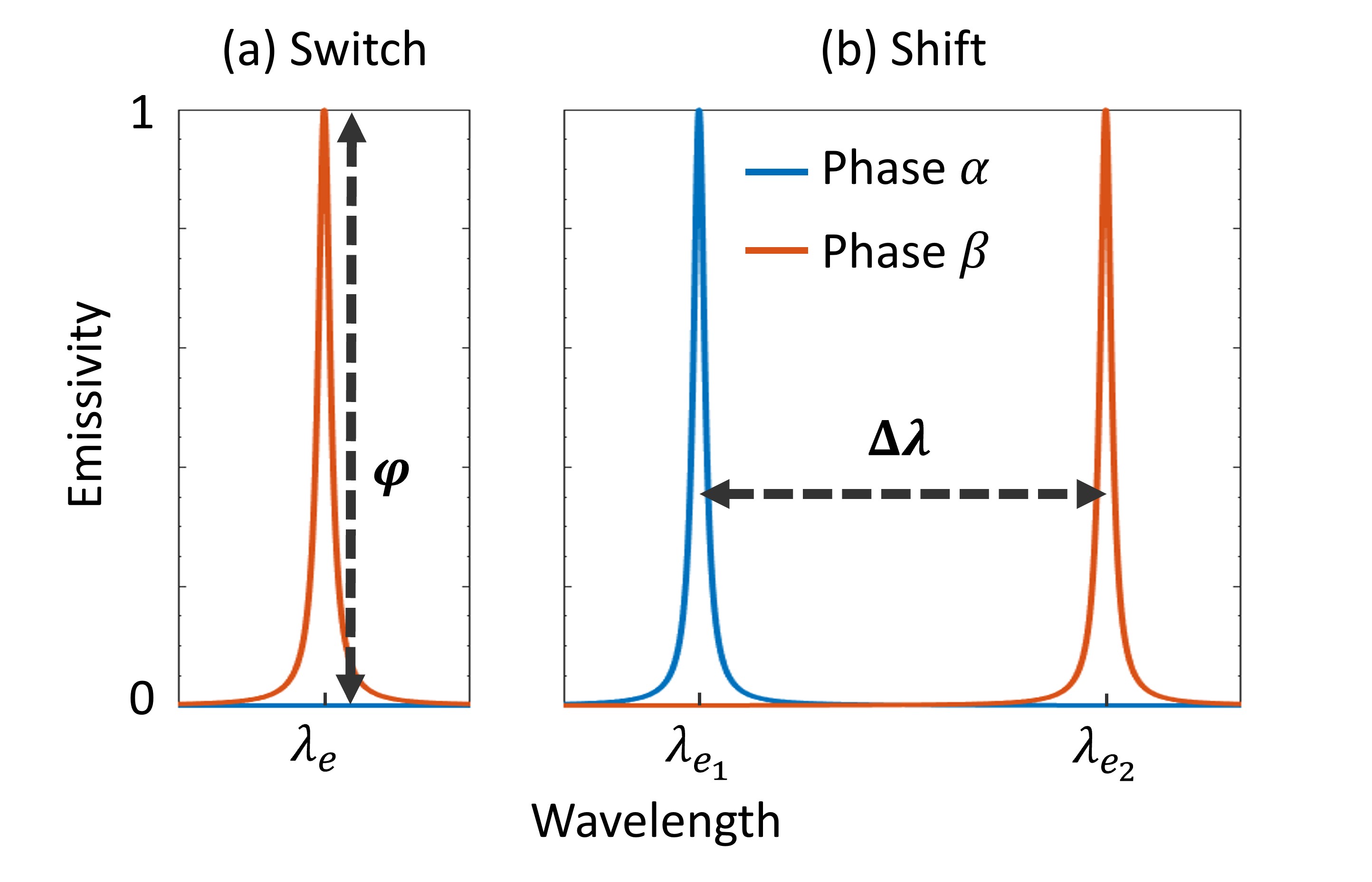}
    \caption{Ideal narrowband spectral emissivity upon the phase transition of a PCM for two configurations: (a) ON-OFF switching and (b) Spectral shifting from one resonance wavelength to another.}
    \label{fig:Figure1}
\end{figure}

In this work, we propose a simple framework combining SPhPs-based resonances and PCMs in a Salisbury screen configuration to design lithography-free narrowband IR emitters with different properties upon phase transition of the PCM. 
We first derive analytical conditions for unitary and zero emissivity. 
We then apply this versatile framework to two configurations. 
In the first, the emissivity of a single-resonance emitter is turned ON and OFF upon phase transition (Fig.~\ref{fig:Figure1}(a)). 
In the second, which considers two arbitrary emission wavelengths, the emission peak switches from one wavelength to the other (Fig.~\ref{fig:Figure1}(b)). 
In both cases, we quantify the performance of optimized devices, considering both idealized and real materials. 
Finally, we show that the emissivity of these structures shows very little angular dependence, making them relevant for spectrally-tunable diffuse narrowband thermal emission. 

\begin{figure}[h]
    \centering
    \includegraphics[width = \columnwidth]{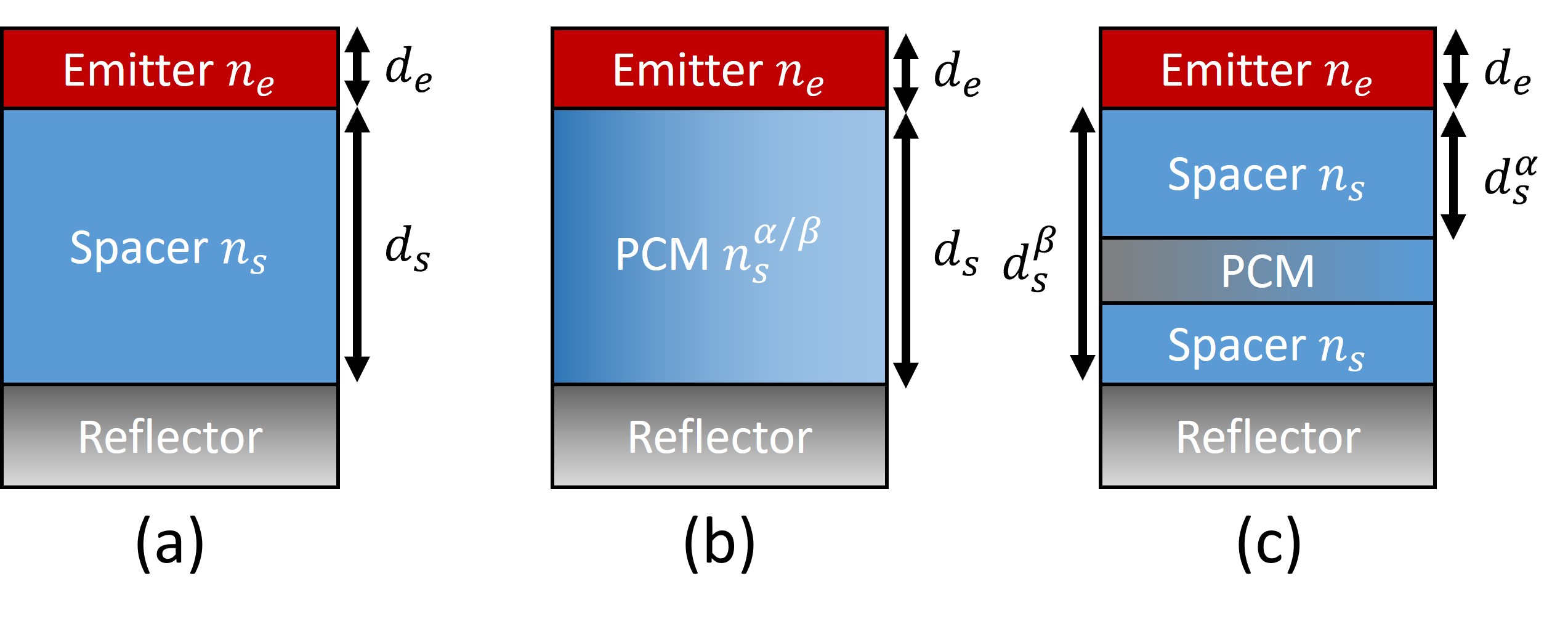}
    \caption{
    (a) Salisbury screen configuration considered for unitary and zero emissivity: a thin emitter is placed on top of a dielectric spacer, above a back reflector.
    (b-c) Two ways to modify the optical thickness of the spacer. (b) When the PCM has a dielectric-to-dielectric transition, it can be used as the spacer. (c) When the PCM has a dielectric-to-metal transition, it can be used as a reflector hiding a second spacer which only plays a role in the dielectric phase.
    }
    \label{fig:Figure2}
\end{figure}

We consider a two-layer stack consisting of an emitter with complex refractive index $n_e$ and thickness $d_e$ on top of a spacer with a real refractive index $n_s$ and thickness $d_s$. 
It is surrounded by two semi-infinite media: an upper medium with real refractive index $n_i$ and a back reflector with a complex refractive index $n_b$. The general architecture is illustrated in Fig.~\ref{fig:Figure2}(a).
The emitter layer supports a SPhP resonance at wavelength $\lambda_e$, and its relative permittivity is $\varepsilon_e = n_e^2$. 
We also define the optical thickness of the spacer as $\Delta_s = n_s d_s$.
We restrict our analysis to normal incidence, and discuss its extension for oblique angles towards the end of the manuscript.
Using Kirchhoff's law of thermal radiation, we describe emissivity as $\epsilon = 1-R$, where $R = |r|^2$ is the reflectivity, $r$ being the Fresnel reflection coefficient of the system.
We consider the incident medium to be air ($n_i = 1$) and the back medium to be a perfect reflector ($1/n_b \to 0$). 
The following results are derived and generalized in the Supplemental Material~\cite{suppl}, starting from ref.~\cite{sarkar_lithography-free_2022}.
Assuming $\Im(\varepsilon_e(\lambda_e)) \gg 1$ ($\Im$ meaning the imaginary part), the emitter thickness required to achieve unitary emissivity (resonance) at wavelength $\lambda_e$ is:

\begin{equation}
    d_e = \frac{\lambda_e}{2 \pi \ \Im(\varepsilon_e(\lambda_e))}, 
    \label{eq:d_e}
\end{equation}

\noindent whereas the spacer's optical thickness is:
    
\begin{equation}
     \Delta_{s}^{\epsilon=1} = \left(m +\frac{1}{2} \right) \frac{\lambda_e}{2}, \ m\in\mathbb{N}, 
     \label{eq:Delta_s_R=0}
\end{equation}

\noindent where $\mathbb{N}$ is the set of all natural numbers, including zero.
On the contrary, imposing the condition of zero emissivity and considering the same emitter thickness (Eq.~\ref{eq:d_e}), the spacer's optical thickness becomes:

\begin{equation}
    \Delta_{s}^{\epsilon=0} = m \frac{\lambda_e}{2}, \ m\in\mathbb{N}
    \label{eq:Delta_s_R=1}
\end{equation}

Therefore, if $\Im(\varepsilon_e) \gg 1$, then $d_e \ll \lambda_e$ (Eq.~\ref{eq:d_e}), hence we can neglect the impact of the front layer on the resonance conditions, leading to the resonance and anti-resonance wavelengths of a simple Fabry-Perot cavity (Eqs.~\ref{eq:Delta_s_R=0}-\ref{eq:Delta_s_R=1}).

The emissivity of such a system can be actively tuned between 0 and 1 by modifying the optical thickness of the spacer, $\Delta_s = n_s d_s$. This can be achieved with PCMs. In particular, a PCM that switches from one dielectric phase to another dielectric phase, with different refractive indices $n_s^{\alpha}$ and $n_s^{\beta}$, can be used directly as the spacer. 
The index change upon phase transition modulates the optical thickness and thus the emissivity of the system (Fig.~\ref{fig:Figure2}(b)). Alternatively, a PCM that switches from a dielectric phase to a metallic phase can be inserted between two spacer layers such that the thickness of the cavity itself changes upon phase transition (Fig.~\ref{fig:Figure2}(c)). 
In its metallic phase, the PCM behaves like a perfect reflector, and the spacer thickness is $d_{s}^{\alpha}$, while in its dielectric phase, the spacer becomes the combination of the two spacers and the PCM, with a total thickness $d_{s}^{\beta}$. Note that for the two-layer model (Fig.~\ref{fig:Figure2}(a)) to be analytically valid, the PCM should either be extremely thin or have the same refractive index as the spacer when the PCM is in its dielectric phase. The model allows for unitary emissivity at the resonance wavelength to be achieved either in phase $\alpha$ or $\beta$ (depending on the thickness of the spacer(s)).

In the following, we illustrate this framework with two examples, the first one pertaining to ON-OFF switching of thermal emission at a certain wavelength upon phase transition, and the second one regarding spectral shifting between two frequencies, where the emission wavelength is arbitrarily tuned. 
In both cases, we present the results obtained with both idealized and real materials.
The simulations are performed using an in-house transfer matrix method~\cite{giteau_resonant_2022}.
The refractive index spectra used for all materials considered are represented in the Supplemental Material~\cite{suppl}. The materials considered in this work, with the exception of the PCMs, show negligible dependence in their permittivity with temperature.

As a first illustration, we design a system with narrowband unitary emissivity in phase $\alpha$ and zero emissivity in phase $\beta$. 
The figure of merit for evaluating the performance of the switch can be defined as the difference in emissivity between the two phases, at the resonance wavelength $\lambda_e$ of the emitter:

\begin{equation}
    \varphi_{switch} = \epsilon^\alpha(\lambda_e)-\epsilon^\beta(\lambda_e).
\end{equation}

For ideal materials, $\varphi_{switch}$ should be unity.
By considering Eqs.~\ref{eq:Delta_s_R=0}-\ref{eq:Delta_s_R=1} in this particular system, we derive the ON and OFF conditions, respectively, for the optical thickness of the PCM spacer:
\begin{align} \label{eq:example1_spacer_alpha}
    \Delta_{s}^{\alpha} &= \left(m_{\alpha} +\frac{1}{2} \right) \frac{\lambda_e}{2}, \ m_{\alpha}\in\mathbb{N}, \\ \label{eq:example1_spacer_beta}
    \Delta_{s}^{\beta} &= m_{\beta} \frac{\lambda_e}{2}, \ m_{\beta}\in\mathbb{N},
\end{align}
\noindent which imposes
\begin{equation}
    \frac{\Delta_{s}^{\alpha}}{\Delta_{s}^{\beta}} = \frac{2 m_{\alpha} + 1}{2 m_{\beta}}.
    \label{eq:ratio_Delta_s_switch}
\end{equation}

Note that, to minimize the thickness of the spacer, the indices $m_{\alpha}$ and $m_{\beta}$ should be as small as possible.

Here, we consider the configuration where the PCM is used as the spacer (Fig.~\ref{fig:Figure2}(b)), as it is the simpler configuration. 
Based on Eq.~\ref{eq:ratio_Delta_s_switch}, the refractive index ratio should correspond to:

\begin{equation}
    \frac{n_{s}^{\alpha}}{n_{s}^{\beta}} = \frac{2 m_{\alpha} + 1}{2 m_{\beta}}.
    \label{eq:ratio_n_s_switch}
\end{equation}

Such a configuration is only constrained by the refractive indices of the PCM.
One promising material for this application is $\mathrm{Ge_2Sb_2Se_4Te_1}$ (GSST), which shows a large refractive index ratio, close to $3/2$, in the IR frequency range between its crystalline phase ($n^s_{\alpha}\approx 4.60$) and its amorphous phase ($n^s_{\beta}\approx 3.19$)~\cite{zhang_broadband_2019}. 
Therefore, from Eq. \ref{eq:ratio_n_s_switch}, we consider $m_{\alpha} = m_{\beta} = 1$.
For the sake of illustration, here, we consider a SiC~\cite{spitzer_infrared_1959,le_gall_experimental_1997} emitter with a resonance wavelength $\lambda_e = 12.6 \ \mathrm{\mu m}$. Its thickness, determined from Eq.~\ref{eq:d_e}, is $d_e = 3.7 \ \mathrm{nm}$. 

\begin{figure}[h]
    \centering
    \includegraphics[width = \columnwidth]{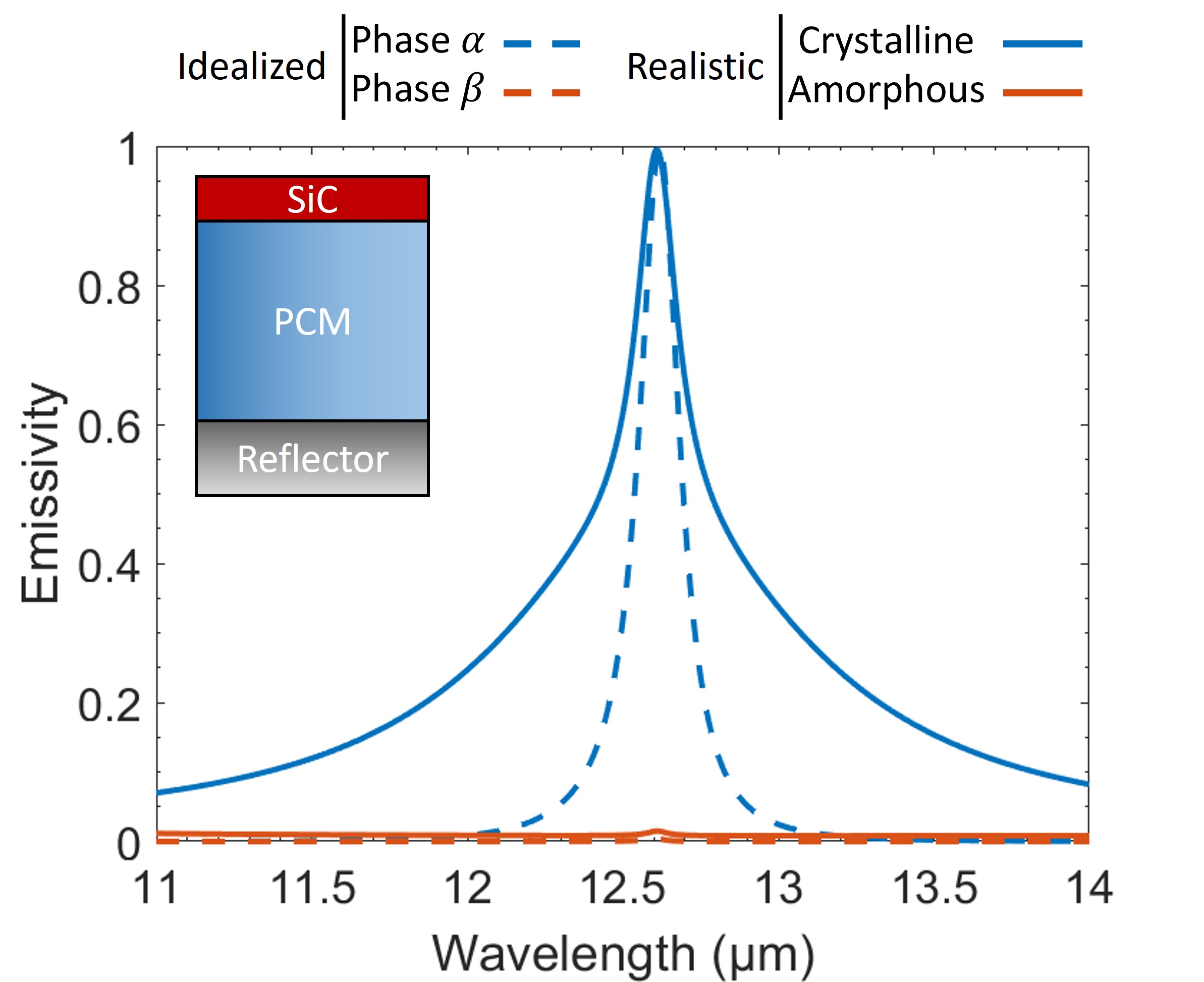}
    \caption{Spectral emissivity for both PCM phases, shown with the blue and red curves, in the ON-OFF switching configuration, using a 3.7 nm-thick SiC emitter with a dielectric-to-dielectric PCM. Dashed line: Assuming an idealized PCM and a perfect back reflector. Solid line: Considering a GSST spacer and a silver back reflector.}
    \label{fig:Figure3}
\end{figure}

For the ideal configuration, we consider a dispersionless and non-absorptive GSST with refractive indices $n_s^{\alpha} = 4.60$ and $n_s^{\beta} = 3.19$ on top of a perfect reflector, so as to satisfy the refractive index ratio of Eq. \ref{eq:ratio_n_s_switch} for $m_{\alpha} = m_{\beta} = 1$. The figure of merit is maximized for a spacer thickness $d_s = 2.015 \ \mathrm{\mu m}$, corresponding to near-ideal ON-OFF switching, with $\varphi_{switch} = 0.996$ (Fig.~\ref{fig:Figure3}).
This spacer thickness is the average of the values predicted by eqs.~\ref{eq:example1_spacer_alpha}-\ref{eq:example1_spacer_beta} ($2.054 \ \mathrm{\mu m}$ and $1.975 \ \mathrm{\mu m}$ , respectively).
The width of the resonance is determined by that of the SPhP mode in SiC.
We note that even with an imperfect refractive index ratio, we can achieve a figure of merit extremely close to unity, demonstrating the imperfection tolerance of the approach.

For a realistic configuration, we consider the complex refractive indices of GSST~\cite{zhang_broadband_2019} and silver~\cite{yang_optical_2015}, for the PCM layer and back reflector, respectively. The PCM thickness is set to $d_s = 2.035 \ \mathrm{\mu m}$ (very close to the ideal case), as it maximizes the figure of merit. 
As can be seen in Fig.~\ref{fig:Figure3}, the ON-OFF switching remains extremely effective, with $\varphi_{switch} = 0.980$.
Nonetheless, we note that emissivity in the crystalline phase is significantly broadened due to parasitic absorption in the PCM and the Ag back reflector (\ref{fig:Figure3}).
The origin of this broadening and strategies to mitigate it are discussed in the Supplemental Material~\cite{suppl}.

The framework discussed above can be extended to several thin emitters stacked on top of each other (or an emitter with several SPhP resonances) to achieve a spectral shift in the emissivity in the two PCM phases. This can be achieved as long as there is no significant spectral overlap between the SPhP modes of the two thin emitters.
Here, we consider a system with two emitters, referred to as $e_1$ and $e_2$ henceforth. These emitters support a SPhP resonance at wavelengths $\lambda_{e_1}$ and wavelength $\lambda_{e_2}$, respectively. The aimed operation is to achieve unitary emission at wavelength $\lambda_{e_1}$ in phase $\alpha$ and unitary emission at wavelength $\lambda_{e_2}$ in phase $\beta$ (Fig.~\ref{fig:Figure1}(b)). 
We define a figure of merit to quantify the quality of the spectral shift, which should approach unity in the ideal case:

\begin{equation}
    \varphi_{shift} = \frac{1}{2} \left[ \left(\epsilon^\alpha(\lambda_{e_1})-\epsilon^\beta(\lambda_{e_1})\right) + \left(\epsilon^\beta(\lambda_{e_2})-\epsilon^\alpha(\lambda_{e_2})\right) \right]
\end{equation}

We assume the spacer is dispersionless, such that $n_s^{\alpha/\beta}(\lambda_{e_1}) = n_s^{\alpha/\beta}(\lambda_{e_2}) = n_s^{\alpha/\beta}$.
Based on Eqs.~\ref{eq:Delta_s_R=0}-\ref{eq:Delta_s_R=1}, the system must satisfy at once the following four equations:

\begin{align}
    \Delta_{s}^{\alpha} &= \left(m_{1} +\frac{1}{2} \right) \frac{\lambda_{e_1}}{2}, \ m_{1}\in\mathbb{N}, \label{eq:ex2_eq1}\\
    \Delta_{s}^{\alpha} &= m_{2} \frac{\lambda_{e_2}}{2}, \ m_{2}\in\mathbb{N}, \label{eq:ex2_eq2} \\
    \Delta_{s}^{\beta} &= \left(m_{3} +\frac{1}{2} \right) \frac{\lambda_{e_2}}{2}, \ m_{3}\in\mathbb{N}, \label{eq:ex2_eq3} \\
    \Delta_{s}^{\beta} &= m_{4} \frac{\lambda_{e_1}}{2}, \ m_{4}\in\mathbb{N}, \label{eq:ex2_eq4}
\end{align}

\noindent which imposes a wavelength ratio:

\begin{equation}
    \frac{\lambda_{e_2}}{\lambda_{e_1}} = \frac{2 m_{1} + 1}{2 m_{2}} = \frac{2 m_{4}}{2 m_{3} + 1}
    \label{eq:lambda_ratio_shift}
\end{equation}

It is impossible to find four integers ($m_{1}$-$m_{4}$) that satisfy Eq.~\ref{eq:lambda_ratio_shift} exactly. Nonetheless, one can achieve a spectral shift close to optimal if the two integer ratios in Eq.~\ref{eq:lambda_ratio_shift} are close to each other.
We emphasize that this approach allows arbitrarily large spectral shifts in emissivity, provided the availability of materials with the desired resonance frequencies, the transparency of the spacer material in that spectral range, and the bandwidth of blackbody radiation. This is in contrast to previous works that rely on tuning the resonance frequency of a single emitter material, therefore generally leading to rather limited frequency shifts~\cite{cao_mid-infrared_2013,xiong_extremely_2022,ma_narrowband_2022}.

To illustrate this case, we consider a system where the PCM has a dielectric-to-metal phase transition (Fig.~\ref{fig:Figure2}(c)).
The advantage of this configuration is that any optical thickness ratio can be achieved by adjusting the relative thickness of the spacers, relaxing the number of constraints.
We consider two emitters: hexagonal boron nitride (hBN), which supports a SPhP mode near $\lambda_{e_1} \approx 7.3 \ \mu m$ ~\cite{cai_infrared_2007,kumar_tunable_2015} and $\rm{\alpha}$-SiC~\cite{spitzer_infrared_1959,le_gall_experimental_1997}, which supports a SPhP mode near $\lambda_{e_2} \approx 12.6 \ \mu m$, with thicknesses $2.3$ nm and $3.7$ nm, respectively (calculated from Eq.~\ref{eq:d_e}).
Such small thicknesses are particularly attractive for low-dimensional layered materials, such as hBN, which can easily be exfoliated into few-monolayer flakes. Nevertheless, the principle of operation of the device does not change for thicker emitter layers, other than slightly shifting the position of the central resonance (Eq.~\ref{eq:d_e}).

\begin{figure}[h]
    \centering
    \includegraphics[width = \columnwidth]{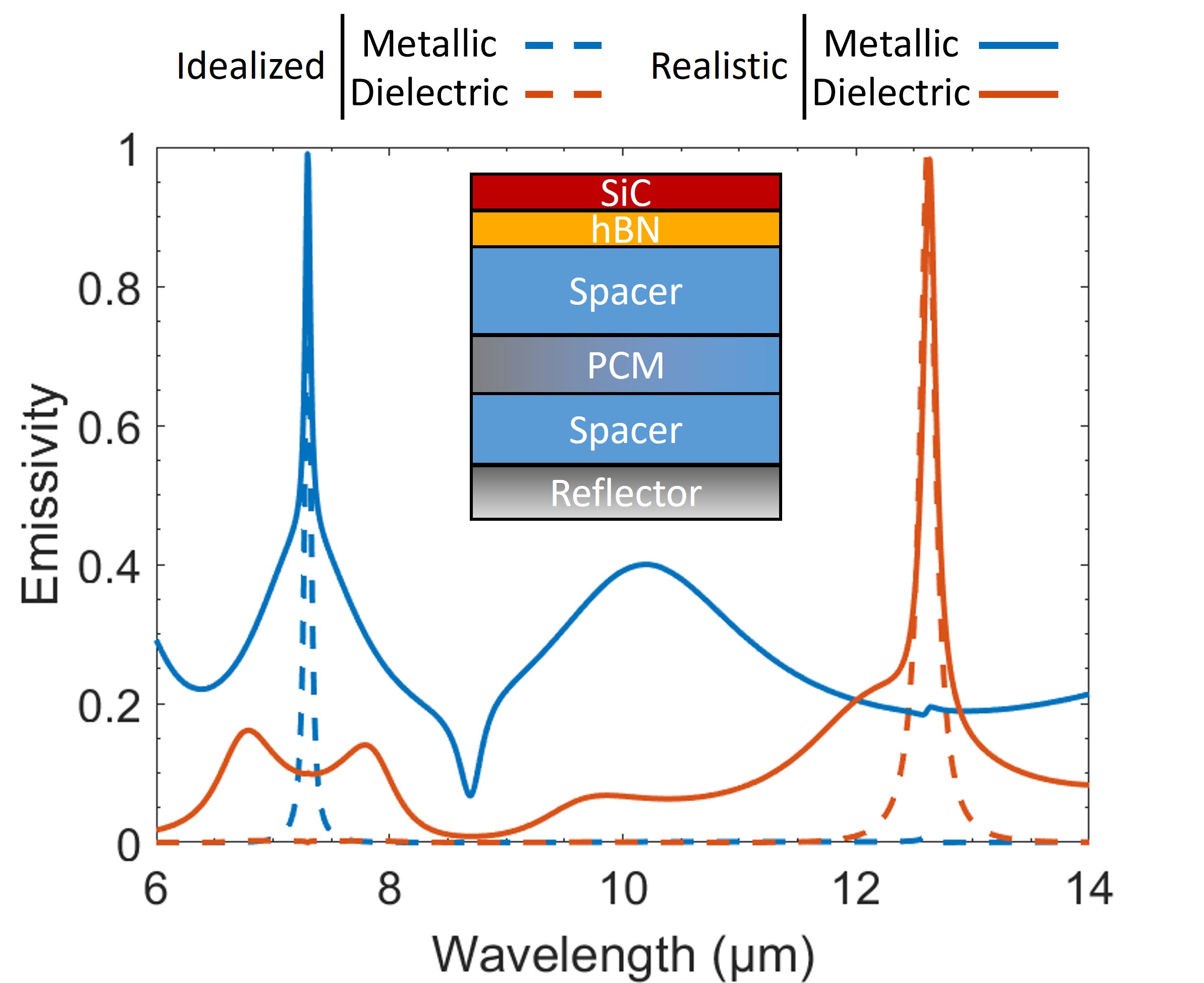}
    \caption{Spectral shifting between two wavelengths using two SPhP emitters and a PCM with a phase transition from metal to dielectric surrounded by two dielectric spacers. Dashed line: Assuming a perfect dielectric-to-metallic PCM and a perfect reflector Solid line: Considering $\mathrm{VO_2}$ as the PCM and a silver back reflector.}
    \label{fig:Figure4}
\end{figure}

The wavelength ratio $\frac{\lambda_{e_2}}{\lambda_{e_1}} \approx 1.73$ can be well approximated by considering the set of integers $m_1 = 3$, $m_2=2$, $m_3=3$ and $m_4=6$. This leads to $\Delta_s^{\alpha} \approx 12.7 \ \mathrm{\mu m}$ and $\Delta_s^{\beta} \approx 22.0 \ \mathrm{\mu m}$.
For the spacer, we consider KBr which is transparent and dispersionless in the spectral region of interest, with $n_s = 1.52$.

In the ideal case, we consider a perfect back reflector. We assume that the PCM has a negligible thickness so that it can be ignored in the dielectric phase, while it behaves like a perfect reflector in the metallic phase (very large refractive index). 
We obtain a near-ideal spectral shift, with a figure of merit $\varphi_{shift} = 0.992$ by considering a top spacer thickness $d_s^\alpha = 8.36 \ \mathrm{\mu m}$ and a bottom spacer thickness of 6.03 µm, for a total spacer thickness $d_s^\beta = 14.4 \ \mathrm{\mu m}$ (Fig.~\ref{fig:Figure4}).
This is consistent with the model predictions $d_s^\alpha \in [8.29 \ \mathrm{\mu m}, 8.40 \ \mathrm{\mu m}]$ and $d_s^\beta \in [14.4 \ \mathrm{\mu m}, 14.5 \ \mathrm{\mu m}]$ (eqs.~\ref{eq:ex2_eq1}-\ref{eq:ex2_eq4}). 

For the realistic configuration, we consider Ag~\cite{yang_optical_2015} as the back reflector, and $\mathrm{VO_2}$ as the PCM~\cite{wan_optical_2019}.
We optimize the thickness of $\mathrm{VO_2}$ as well as that of both spacers to maximize the figure of merit, leading to $\varphi_{shift} = 0.840$ with a top spacer thickness of 8.30 µm, a bottom spacer thickness of 5.62 µm and a $\mathrm{VO_2}$ thickness of 170 nm.
Compared to the ideal case, the $\mathrm{VO_2}$ layer also contributes to the optical path in the dielectric phase, explaining the lower thickness of the bottom spacer.
The $\mathrm{VO_2}$ thickness thus obtained roughly corresponds to its skin depth, ensuring a sufficient reflectivity in the metallic phase while minimizing parasitic absorption in the dielectric phase.
Such $\mathrm{VO_2}$ thicknesses have been deposited on various materials including Si~\cite{morsy_experimental_2020,wan_switchable_2022}.

Compared to the ideal case, $\mathrm{VO_2}$ leads to significant parasitic emission, as can be seen in Fig.~\ref{fig:Figure4}. 
In its metallic phase, $\mathrm{VO_2}$ is significantly less reflective than noble metals, thus broadening the emissivity. 
In its dielectric phase, the imaginary part of the refractive index is not negligible, and increases significantly above 10 µm, leading to parasitic emission.
Using materials with a more drastic dielectric-to-metallic transition, such as $\mathrm{In_3SbTe_2}$~\cite{hesler_in3sbte2_2021}, may enable a behavior closer to the idealized case.

We comment here on the angular dependence of thermal emission for these architectures.
The effective optical thickness of the spacer depends on the angle $\theta$ with respect to normal incidence, following~\cite{yariv_optical_1991}:

\begin{equation}
    \Delta_s(\theta) = d_s \sqrt{n_s^2-\sin^2(\theta)}.
    \label{eq:theta_dependence}
\end{equation}

Therefore, if the refractive index of the spacer is large ($n_s \gtrsim 2$), the optical thickness will depend weakly on the angle. 
As a result, the emission should be diffuse (assuming the angular dependence of front reflectivity plays a minor role). 
Indeed, we confirm in the Supplemental Material~\cite{suppl} that the behavior of the realistic ON-OFF switching structure (Fig.~\ref{fig:Figure3}) is essentially angle-independent up to 40 degrees, owing to the large refractive index of GSST.
As a result, this structure offers a diffuse narrowband IR source that can be turned ON and OFF through control of the PCM phase.

Finally, we discuss the resilience of the system to thickness variations. The emitters considered in this work have a very high quality factor, enabling narrowband thermal emission and leading to very thin optimal thicknesses. Since hBN is a 2D-layered material with a c-axis lattice constant of 0.67 nm~\cite{lynch_effect_1966}, a 4-monolayer emitter would have a thickness of 2.7 nm. Compared to the optimal thickness of 2.3 nm, this makes no difference in device performance. For SiC, the peak absorptivity starts decreasing for emitters thicker than 5 nm, which might be challenging to achieve homogeneously via chemical or physical vapor deposition. Considering emitters with lower quality factors such as $\mathrm{SiO_2}$ can yield to more practical thicknesses around 100 nanometers, at the expense of emission bandwidth. Regarding the spacer thickness, the tolerance is in the order of 50 nm for each layer to maintain optimal performance. Although small, this is within the precision achievable using carefully calibrated thermal evaporation or sputtering.

We have proposed a simple framework to design lithography-free narrowband IR thermal sources which can change their behavior upon the phase transition of a PCM layer. 
We suggested different implementations depending on the nature of the phase transition. We applied it first to ON-OFF emission switching with a single emitter, then to spectral shifting with two emitters, only one emitting in each phase. With ideal, non-absorptive materials, it is easy to achieve near-ideal performance. 
Devices for ON-OFF switching with very high contrast at all angles should currently be manufacturable, although the relatively thick GSST layers required make it challenging to ensure complete and reversible phase transition~\cite{zhang_broadband_2019}.
More complex functions will require further developments in materials to operate at peak performance, mostly to eliminate the parasitic absorption in the PCM.
Additional degrees of control can be considered in order to improve the response and enable new functionalities, for instance by considering dispersive spacers or by adjusting the thickness of the emitters and their position within the stack~\cite{giteau_resonant_2022}.
The switch configuration is particularly promising for thermophotovoltaic systems, enabling on-demand energy generation through control of the PCM phase. In addition, the demonstrated diffuse thermal emission behavior is attractive for optimizing thermophotovoltaic energy conversion~\cite{pfiester_selective_2017,wang_selective_2022}. More generally, these architectures and the associated formalism should be valuable not only for thermal emission applications including infrared sources and thermal camouflage but more broadly for mid-IR photonic devices such as multi-spectral photodetectors.

\section*{acknowledgments}
The authors declare no competing financial interest. G. T. P. acknowledges funding from ”la Caixa” Foundation (ID 100010434), from the PID2021-125441OA-I00 project funded by MCIN /AEI /10.13039/501100011033 / FEDER, UE, from the TED2021-129841A-I00 funded by MCIN/AEI/ 10.13039/501100011033 and by the European Union “NextGenerationEU”/PRTR, and from the European Union’s Horizon 2020 research and innovation programme under the Marie Sklodowska-Curie grant agreement No 847648. The fellowship code is LCF/BQ/PI21/11830019. M.G. acknowledges financial support from the Severo Ochoa Excellence Fellowship. M.E. acknowledges ayuda PRE2020-094401 financiada por MCIN/AEI/ 10.13039/501100011033 y  FSE "El FSE invierte en tu futuro". 
This work is part of the R\&D project CEX2019-000910-S, funded by MCIN/ AEI/10.13039/501100011033/, from Fundació Cellex, Fundació Mir-Puig, and from Generalitat de Catalunya through the CERCA program.

\bibliography{references.bib}

\end{document}


\title{Design rules for active control of narrowband thermal emission using phase-change materials -- \\ Supplemental Material}

\author{Maxime Giteau}
\author{Mitradeep Sarkar}
\author{Maria Paula Ayala}
\author{Michael T. Enders}
\author{Georgia T. Papadakis}
 \email{georgia.papadakis@icfo.eu}
\affiliation{ICFO-Institut de Ciencies Fotoniques, The Barcelona Institute of Science and Technology, Castelldefels (Barcelona) 08860, Spain}

\maketitle

\begin{figure*}[t]
    \centering
    \includegraphics[width = 160mm]{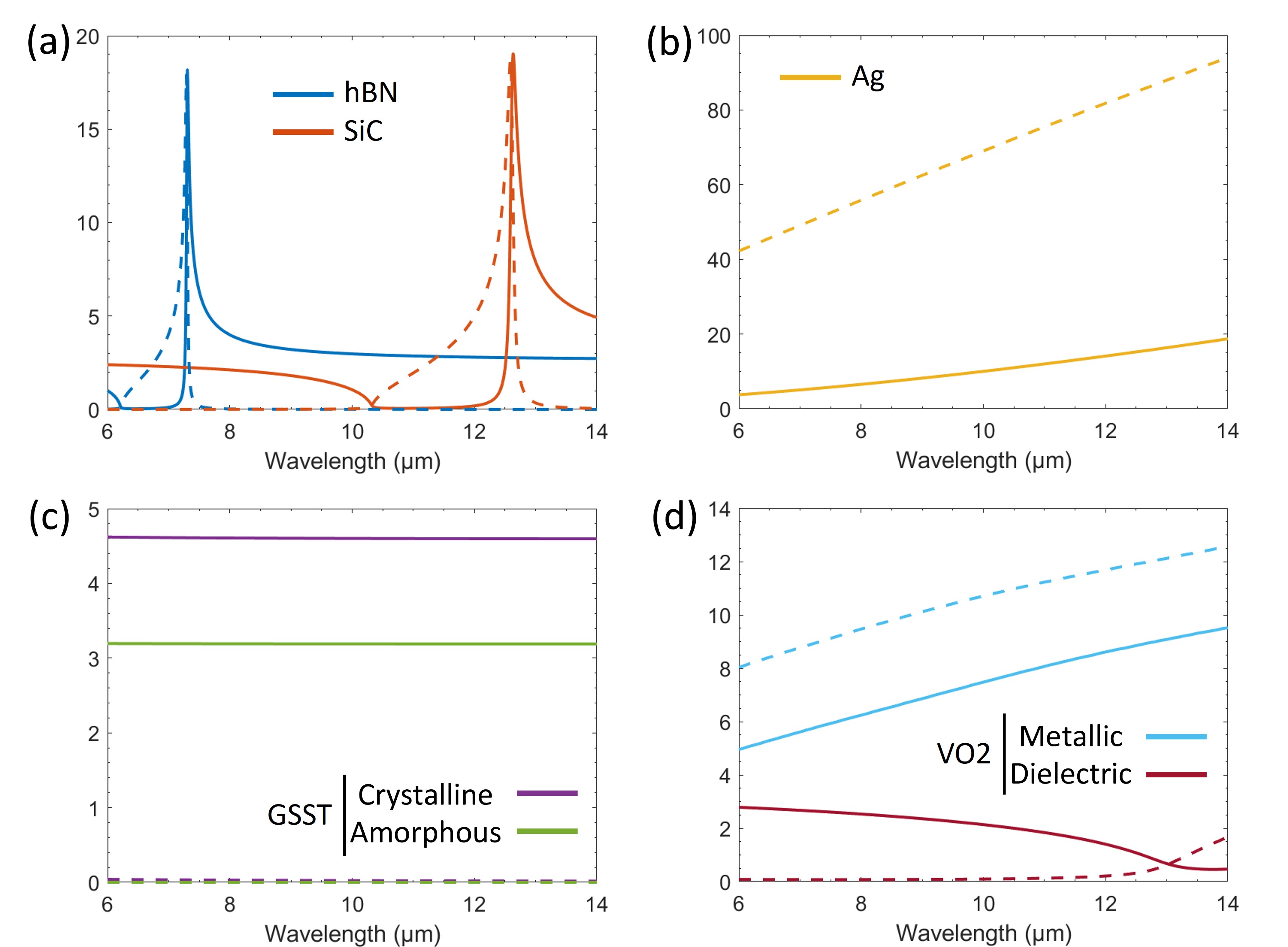}
    \caption{Real part (continuours line) and imaginary part (dashed line) of the refractive index for the materials considered in the paper, as a function of the wavelength.}
    \label{fig:FigureS3}
\end{figure*}

\section{Conditions for unitary and zero emissivity}
\label{sec:conditions}

For normal incidence, the Fresnel reflection coefficient of a 2-layer system takes the expression~\cite{sarkar_lithography-free_2022}:

\begin{widetext}
\begin{equation}
    r = \frac{\left[\left(1-\frac{n_i}{n_b}\right) + \left(\frac{n_i n_s}{n_b n_e} -\frac{n_e}{n_s} \right) T_e T_s \right] 
    + i \left[\left(\frac{n_i}{n_e}-\frac{n_e}{n_b}\right) T_e + \left(\frac{n_i}{n_s} -\frac{n_s}{n_b} \right) T_s \right]}
    {\left[\left(1+\frac{n_i}{n_b}\right) - \left(\frac{n_i n_s}{n_b n_e} +\frac{n_e}{n_s} \right) T_e T_s \right] 
    - i \left[\left(\frac{n_i}{n_e}+\frac{n_e}{n_b}\right) T_e + \left(\frac{n_i}{n_s} +\frac{n_s}{n_b} \right) T_s \right]},
\end{equation}
\end{widetext}

\noindent with $T_{s/e} = \tan(k_0 n_{s/e} d_{s/e})$, where $k_0 = 2\pi/\lambda$ is the wavevector in vacuum of a monochromatic plane wave with wavelength $\lambda$. 
If we consider the incident medium is air ($n_i = 1$) and a perfect back reflector ($1/n_b \to 0$), then the reflection coefficient can be written as:

\begin{equation}
    r = \frac{n_s-n_e T_e T_s + i \left( \frac{n_s}{n_e}T_e +T_s \right)}{n_s-n_e T_e T_s - i \left( \frac{n_s}{n_e}T_e +T_s \right)},
    \label{eq:r_normal}
\end{equation}

\noindent with $T_{s/e} = \tan(k_0 \ n_{s/e} \ d_{s/e})$, where $k_0 = 2\pi/\lambda$ is the wavevector in vacuum of a monochromatic plane wave with wavelength $\lambda$.

\subsection{Unitary emission}

Unitary emission requires $r=0$, implying that both the real and imaginary parts of $r$ should be 0. 
From Eq.~\ref{eq:r_normal}, this corresponds to:

\begin{align}
    n_s &= T_s \Re(n_e T_e) + n_s \Im\left( \frac{T_e}{n_e}\right) \\
    T_s \Im(n_e T_e) &= n_s \Re\left( \frac{T_e}{n_e}\right) + T_s.
\end{align}

If there are no further assumptions, this system should be solved directly to determine the optimal thickness of both the emitter and the spacer. 
Assuming $T_e \approx k_o n_e d_e$, the system simplifies into:
\begin{align}
    n_s &= T_s k_0 d_e \Re(\varepsilon_e) \label{eq:cond1}\\
    T_s k_0 d_e \Im(\varepsilon_e) &= n_s k_0 d_e + T_s. \label{eq:cond2}
\end{align}

The condition on the real part (Eq.\ref{eq:cond1}) gives:

\begin{equation}
    T_s = \frac{n_s}{k_0 d_e \Re(\varepsilon_e)},
    \label{eq:T_s_vs_epsilon_e}
\end{equation}

\noindent which we can inject into Eq.~\ref{eq:cond2} to obtain the quadratic equation:

\begin{equation}
    \Re(\varepsilon_e) X^2 -\Im(\varepsilon_e) X + 1 = 0,
\end{equation}

\noindent where $X=k_0 d_e$. Eq.~\ref{eq:T_s_vs_epsilon_e} admits 2 solutions, of the form:

\begin{equation}
    X = \frac{\Im(\varepsilon_e)}{2 \Re(\varepsilon_e)} \left[ 1\pm \sqrt{1-\frac{4\Re(\varepsilon_e)}{\left[\Im(\varepsilon_e)\right]^2}} \right].
\end{equation}

Assuming $\frac{4\Re(\varepsilon_e)}{\Im(\varepsilon_e)^2} \ll 1$, only the solution with the minus sign is compatible with the assumptions. 
It simplifies to:

\begin{equation}
    X = \frac{1}{\Im(\varepsilon_e)},
\end{equation}

\noindent which gives directly Eq.~1 of the manuscript.
Combining this solution with Eq.~\ref{eq:T_s_vs_epsilon_e}, we find:

\begin{equation}
    \Delta_s = \frac{\lambda}{2 \pi} \left[ \arctan\left(\frac{n_s \Im(\varepsilon_e)}{\Re(\varepsilon_e)}\right) +m\pi \right], \quad m \in \mathbb{N}.
\end{equation}

In the limit where $n_s \Im(\varepsilon_e) \gg \Re(\varepsilon_e)$, the arctan term tends to $\pi/2$, leading to Eq.~2 in the manuscript.

\subsection{Zero emissivity}

Emissivity is zero when the reflectivity $|r|^2 = 1$. 
From Eq.~\ref{eq:r_normal}, a sufficient condition for unitary reflection is to cancel the term in the parentheses, corresponding to:

\begin{equation}
    \frac{T_s}{n_s} = -\frac{T_e}{n_e}.
\end{equation}

Using the linear approximation of $T_e$ and the emitter thickness obtained for unitary emission, we get:

\begin{equation}
    T_s = -\frac{n_s}{\Im(\varepsilon_e)},
\end{equation}

\noindent or equivalently

\begin{equation}
    \Delta_s = \frac{\lambda}{2 \pi} \left[ -\arctan\left(\frac{n_s }{\Im(\varepsilon_e)}\right) +m\pi \right], \quad m \in \mathbb{N^*}.
\end{equation}

Considering $n_s \ll \Im(\varepsilon_e)$, we obtain, in the zeroth-order approximation, Eq.~3 of the manuscript.

\section{Refractive indices}

The refractive index spectra for all materials considered in this work are represented in Fig.~\ref{fig:FigureS3}. As can be seen in panel (a), hBN and SiC support phonon resonances at wavelengths of 7.3 µm and 12.6 µm, respectively. Their data is taken from refs.~\cite{cai_infrared_2007} and \cite{spitzer_infrared_1959}, respectively. Ag is modelled via the Drude model, and its permittivity, taken from ref.~\cite{yang_optical_2015}, is shown in panel (b). For GSST, we considered the data from ref.~\cite{zhang_broadband_2019}, shown in panel (c). Similarly, the optical properties of $\mathrm{VO_2}$ were taken from ref.~\cite{wan_optical_2019} (film 3) and are shown in panel (d).

\section{Reducing resonance broadening}
\label{sec:broadening}

\begin{figure*}[!htb]
    \centering
    \includegraphics[width = 160 mm]{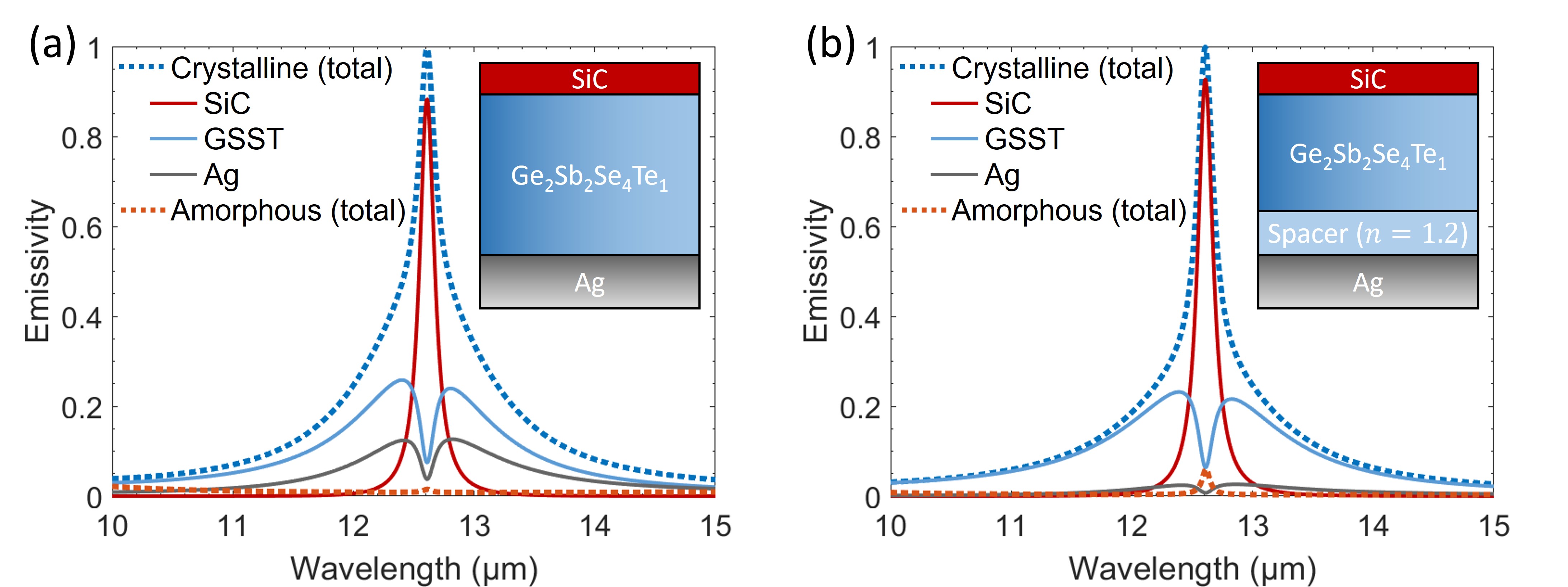}
    \caption{(a) Spectral emissivity for both PCM phases for the ON-OFF switching of a 3.7 nm-thick SiC emitter illustrated in Fig.~3 of the manuscript, along with the contribution to emissivity from each layer.
    (b) The same architecture with a low-index ($n = 1.2$) 1 µm-thick dielectric spacer (and a GSST thickness reduced to 1.53µm) inserted directly above the Ag back reflector. The contribution from Ag is suppressed and that of GSST is reduced, while the emissivity in the amorphous phase increases slightly.}
    \label{fig:FigureS1}
\end{figure*}

\begin{figure*}[!htb]
    \centering
    \includegraphics[width = 160mm]{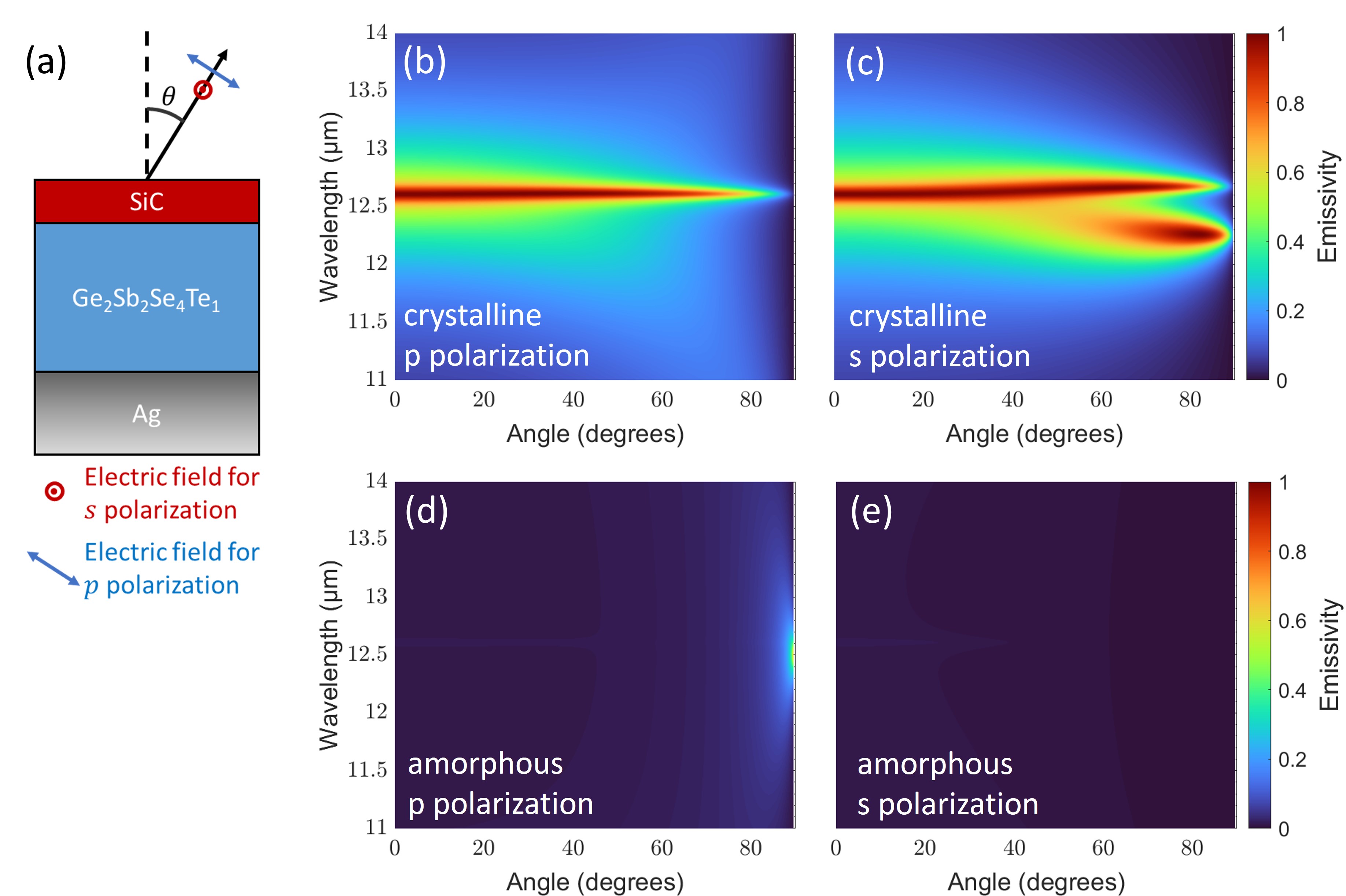}
    \caption{Emissivity as a function of the emission angle and wavelength in the realistic ON-OFF switching example introduced in Fig.~3 of the manuscript. (a) Illustration of the system, including the definition of the angle of incidence and s and p polarizations (b,c) In the crystalline phase (unitary emissivity), for p and s polarizations, respectively. (d,e) In the amorphous phase (zero emissivity), for p and s polarizations, respectively.}
    \label{fig:FigureS2}
\end{figure*}

When considering real materials in the emission switching example, the emissivity is broadened significantly due to parasitic absorption (Fig.~3 in the manuscript). 
Computing the absorption in each layer, we find that about 1/3 of this parasitic emission originates from Ag, and the rest from GaAs (Fig.~\ref{fig:FigureS1}(a)).
(Note that the absorption in GSST might be unreliable since the measurement accuracy for the imaginary part of the refractive index $\kappa$ in Ref.~\cite{zhang_broadband_2019} is about 0.02, while the data gives $\kappa \approx 0.018$ around 12.6 µm.)
We consider adding a low-index dielectric spacer ($n=1.2$ and thickness 1 µm) below the PCM (with a reduced thickness of 1.53 µm) to decrease emission broadening (Fig.~\ref{fig:FigureS1}(b)).
We observe a significant reduction of the back reflector's absorption thanks to a lower electric field intensity at the Ag surface, while GSST absorption is also slightly decreased thanks to the thickness reduction. 
This means, however, that the optical thickness ratio between both PCM phases is no longer optimal, leading to non-zero emissivity in the amorphous phase, and thereby a slight decrease in the figure of merit, with $\varphi_{switch} = 0.939$.

\section{Angular dependence: diffuse emitters}
\label{sec:angle}

We show in Fig.~\ref{fig:FigureS2}, for both polarizations, the angular and spectral dependence of emissivity for the realistic ON-OFF switching structure introduced in Fig.~3 of the manuscript.
In the amorphous phase, the emissivity is close to zero for all angles, wavelengths, and polarizations. 
In the crystalline phase, the emission is narrowband and diffuse for both polarizations (the second peak for high angles in s polarization originates from the parasitic absorption in GSST and Ag). These features can be attributed to the large refractive index of GSST in both phases (Eq.~15 in the manuscript).
